\documentclass[a4paper]{spie}  

\usepackage{amsmath,amsfonts,amssymb}
\usepackage{graphicx}
\usepackage[colorlinks=true, allcolors=blue]{hyperref}

\title{VIALACTEA knowledge base\\homogenizing access to Milky Way data}

\author[a]{Marco Molinaro}
\author[a]{Robert Butora}
\author[b]{Marilena Bandieramonte}
\author[b]{Ugo Becciani}
\author[c]{Massimo Brescia}
\author[c]{Stefano Cavuoti}
\author[b]{Alessandro Costa}
\author[d]{Anna M. Di Giorgio}
\author[d]{Davide Elia}
\author[e]{Akos Hajnal}
\author[e]{Hermann Gabor}
\author[e]{Peter Kacsuk}
\author[d]{Scige J. Liu}
\author[d]{Sergio Molinari}
\author[c]{Giuseppe Riccio}
\author[d]{Eugenio Schisano}
\author[b]{Eva Sciacca}
\author[a]{Riccardo Smareglia}
\author[b]{Fabio Vitello}

\affil[a]{INAF - Osservatorio Astronomico di Trieste, via G.B.~Tiepolo 11 -
34143 Trieste, Italy}
\affil[b]{INAF - Osservatorio Astrofisico di Catania, Via S.~Sofia 78 - 95123
Catania, Italy}
\affil[c]{INAF - Osservatorio Astronomico di Capodimonte, Salita Moiariello 16 -
80131 Napoli, Italy}
\affil[d]{INAF - Istituto di Astrofisica e Planetologia Spaziali, Area di
Ricerca di Tor Vergata,
via Fosso del Cavaliere 100 - 00133 Roma, Italy}
\affil[e]{MTA-SZTAKI, 1111 Budapest, Kende u. 13-17, Hungary}

\authorinfo{Further author information: (Send correspondence to Marco
Molinaro)\\Marco Molinaro: E-mail: molinaro@oats.inaf.it, Telephone: +39 040
3199 152}

\begin{document} 
\maketitle

\begin{abstract}
The VIALACTEA project has a work package dedicated to ``Tools and
Infrastructure'' and, inside it, a task for the ``Database and Virtual
Observatory Infrastructure''. This task aims at providing an infrastructure to
store all the resources needed by the, more purposely, scientific work packages
of the project itself. This infrastructure includes a combination of: storage
facilities, relational databases and web services on top of them, and has taken,
as a whole, the name of VIALACTEA Knowledge Base (VLKB). This contribution
illustrates the current status of this VLKB. It details the set of data
resources put together; describes the database that allows data discovery
through VO inspired metadata mainteinance; illustrates the discovery, cutout and
access services built on top of the former two for the users to exploit the data
content.
\end{abstract}

\keywords{Databases, VO, Data Access, Milky Way}

\section{INTRODUCTION}
\label{sec:intro}
The VIALACTEA project goal is to bring to a common forum the major
new-generation surveys of the Galactic Plane in the radio band, both in thermal
continuum and in atomic and molecular lines, from Europe-funded space missions
and ground-based facilities, to engage one of the fundamental challenges in
Galactic astronomy: to quantify in our galaxy
the relationship between the physical agents responsible for the onset and the
regulation of star formation
in a spiral galaxy and the resulting rate and efficiency of star formation, and
obtain a ``star formation recipe'' that
will be a cornerstone to trace the star formation history of galaxies back to
their formation.

Among the work packages the project is composed of, one incorporates the
activities needed to ensure that a computing infrastructure and a set of
data-mining, machine-learning, visual analytics and 3D visualization tools are 
in place according to astronomers specifications.

The input data that the computational tasks should start from, as well as the output
results the various tasks will provide form a knowledge base that is, in turn, the 
goal of one task of this latter work package.

This contribution focuses on this task, the creation and maintenance, as well as the
access interfaces, of the VIALACTEA Knowledge Base (VLKB).

Section \ref{sec:vialactea} will briefly describe the project work packages and tasks and
give a more detailed view of the ``tools and infrastructure'' work package (Sec.
\ref{subsec:wp5}), with specific attention to the VLKB (Sec.~\ref{subsubsec:wp5task1}).
Section \ref{sec:collect} will then focus on the heterogeneous content of the storage 
(Sec.~\ref{subsec:storage}) available to the VIALACTEA community and the database 
(Sec.~\ref{subsec:database}) that contains the storage metadata descriptions as well as 
catalogued data consumed or provided by the project's scientific tasks.
Next section (Sec.~\ref{sec:interfaces}) will then describe the search and access 
interfaces that live on top of the above data collections; Sec.~\ref{subsec:tap} focuses 
on the generic TAP interface (Table Access Protocol\cite{2010ivoa.spec.0327D}~, an IVOA standard) 
that exposes the database content in an interoperable way while Sec.~\ref{subsec:dcm} describes 
in details the dedicated discovery and access interfaces that allow filtering of data 
collections and retrieval of the actual data files.

\section{PROJECT OVERVIEW}
\label{sec:vialactea}
The VIALACTEA project consists of four scientific work packages (WPs), a technical one 
(WP 5, that we will touch in more details in Sec.~\ref{subsec:wp5}) and other packages 
dedicated to dissemination, management and coordination activities.
Scientific packages' tasks span from diffuse and compact structure analysis (WP 1 \& 2) to 
distance estimation (WP 3) and their outputs are then combined into a global scenario of 
the galaxy as a star formation engine within WP 4 tasks.
All the scientific tasks need some data to start from and produce some output to be 
preserved: 
\begin{itemize}
	\item diffuse structure analysis produce catalogues of filamentary and bubble 
structure out of continuum radio surveys using morphological analysis and radiative 
transfer modeling;
	\item compact source analysis provides a catalogue of compact sources out of 
continuum surveys and combines it with catalogues at different wavelengths to have a multi 
band catalogue to fit against synthetic protocluster models (another project output);
	\item distance estimation uses kinematical analysis of existing radio spectroscopic 
	surveys and project generated 3D extinction maps to produce distance estimation of 
	compact sources or other diffuse regions of the galaxy plane.
\end{itemize}
All of these tasks output are then combined into WP4 tasks for the final scientific goals 
of the project.

The goal of WP 5, ``Tools and Infrastructures'' is to provide technical support in running 
and developing the tools and recipes needed by the scientific counterpart.

\subsection{VIALACTEA Tools and Infrastructures}
\label{subsec:wp5}
There are four main activities in this work package:
\begin{description}
	\item[Database and Virtual Observatory Infrastructure:] to have a common set of 
	resources to be used for scientific input and collect the project outputs. This is what 
	this paper is for, the main product of this activity being the setup and maintenance of 
	the VLKB and its interfaces;
	\item[Data Mining Systems:] to develop intelligent integrated systems directly 
	supporting scientific decision making and situation awareness by dynamically
integrating, correlating, fusing and analysing extremely large volumes of disparate data 
resources and streams;
	\item[3D visual analytics systems:] allowing the astronomer to easily conduct research 
	activities using advanced visual methods for multidimensional data and information 
	visualization, real-time data interaction to carry out complex tasks for multi-criteria 
	data/metadata queries for subsample selection and further analysis, or real-time
	control of data fitting to theoretical models;
	\item[Science Gateway:] to enable workflows based on the analysis tools and recipes and 
	run them through the WS-PGRADE/gUSE gateway system.
\end{description}
The tools and infrastructure developed through these activities should bring the 
researcher from the input data to the result in an easy reproducible way, moving, at least 
partially, the steps of data retrieval, decision making, visual inspection of the 
information and/or results and job runs from manual tasks to automated ones.

The VLKB acts in this scenario as the data resource repository with interoperable 
capabilities and custom access interfaces. As already mentioned, the resources collection 
will be reported in Sec.~\ref{sec:collect} and the interfaces on top of them in the 
subsequent Sec.~\ref{sec:interfaces}.

\subsubsection{Database and Virtual Observatory Infrastructure}
\label{subsubsec:wp5task1}
The task activity within which the VLKB has been developed was defined as the one to carry 
out the archival goals of the project as well as the data repository for all of the 
initial data resources needed by the scientific work packages.

The idea was to have a common resource to follow the VO guidelines for interoperability 
while paying specific attention to the project needs. One of the main issues in keeping 
the pace with the IVOA standards has been to develop a project on the most recent 
specifications the VO was discussing upon. This would have been of great help in the 
VIALACTEA scenario if some specific internal requirements were not in conflict. The result 
is that the VLKB and its interfaces are currently not completely developed using IVOA 
protocols, but are ready to be mapped onto them given some time resource (which may be 
tricky since VIALACTEA is at the end of its life cycle). 

The actual result of this task activity (which will be better summarized in Sec.~
\ref{sec:status}) is a resource composed of a set of heterogeneous data collections of 
observational data, in form of images or multi-dimensional datasets and catalogues, 
alongside database schemata and relations dedicated both to metadescription of the 
mentioned data collections and catalogues produced by the project work package activities.

The discovery, access and retrieval solutions on top of these common resource (the VLKB) is based on an IVOA TAP service for all of the database content that needs to be exposed to the community plus some dedicated search, cutout and merge solutions for the 2D and 3D datasets available through the project. This latter secures the datasets (which are a mix of public and private policy ones) and allows search also for project specific datasets, like the 3D extinction maps of the galaxy produced inside the project itself.

\section{VLKB DATA COLLECTIONS}
\label{sec:collect}
The data content of the VLKB comes in two main archival formats:
\begin{itemize}
	\item storage FITS files;
	\item relational database schemata.
\end{itemize}
In this section we describe them both as are currently available (or planned by the end of 
the project) following the above distinction, instead of following an input vs. output 
categorization or a task-based one. Descriptions will take into account:
\begin{itemize}
	\item dataset origin: from within the project or mirror/redistribution of an existing resource;
	\item resource weight (in terms of datasets, file sizes, catalogue records);
\end{itemize}
The interfaces to discovery, access and retrieval of the resources is left to the following 
section (Sec.~\ref{sec:interfaces}).

\subsection{Storage Content}
\label{subsec:storage}
The storage contains files in FITS\cite{2010AA...524A..42P} format only, but they're 
rather heterogeneous. Those FITS files, indeed, span from 2D images in the radio continuum 
to 3D FITS cubes containing radio velocity spectra at specific molecular lines and also a 
collection of 3D extinction maps.

But the number of dimensions and type of axis is not the only difference between the 
various data collections deployed. There are differences in coordinate system and sky 
frame references, galactic coordinates versus equatorial ones to explicit the most easy 
one to handle, as well as degenerate axis references and multiple HDUs or non-standard 
keywords in the FITS header.

Most prominent, however, were the actual observational differences, that is, the various 
surveys, and observational bands within them, that have been used in the scientific 
analysis of the VIALACTEA project. Taking into accout what has been labeled as a 
\textit{sub-survey} (i.e. a collection of data from a specific survey or pointed archive 
referring to only one single molecular or band or other specific metadata), about 50 
different data collections have been put together to be searched and accessed as a unique 
resource.

Apart from the Hi-Gal sub-surveys, the core resource from which most of the VIALACTEA 
primary products are derived, all of the other resources were retrieved from already 
public repositories or released to the VIALACTEA community but retaining their private 
data policy. That's the reason why access to the VLKB infrastructure has been secured and 
allowed only to the project's members.

Tables~\ref{tab:cubes}, \ref{tab:images} and \ref{tab:extmaps} summarize the sub-surveys data collections reporting minimal figures for the various FITS file sets.

\begin{table}[!ht]
\caption{VIALACTEA VLKB stored surveys consisting of FITS data cubes.} 
\label{tab:cubes}
\begin{center}       
\begin{tabular}{|l|l|r|r||l|l|r|r|}
\hline
\rule[-1ex]{0pt}{3.5ex} \textbf{Name} & \textbf{sub-survey} & \textbf{\# files} & \textbf{size [GB]} &
\textbf{Name} & \textbf{sub-survey} & \textbf{\# files} & \textbf{size [GB]} \\ \hline
\rule[-1ex]{0pt}{3.5ex} MOPRA & 12CO & 52 & 45  & MALT90 & HCO+ & 2012 & 23  \\ \hline
\rule[-1ex]{0pt}{3.5ex} MOPRA & 13CO & 52 & 30  & MALT90 & HCN & 2012 & 23  \\ \hline
\rule[-1ex]{0pt}{3.5ex} MOPRA & C17O & 51 & 14  & MALT90 & N2H+ & 2012 & 23  \\ \hline
\rule[-1ex]{0pt}{3.5ex} MOPRA & C18O & 51 & 24  & MALT90 & HNC & 2012 & 23  \\ \hline
\rule[-1ex]{0pt}{3.5ex} CHIMPS & 13CO & 224 & 18  & MALT90 & 13C34N & 2012 & 23  \\ \hline
\rule[-1ex]{0pt}{3.5ex} CHIMPS & C18O & 223 & 20  & MALT90 & 13CS & 2012 & 23  \\ \hline
\rule[-1ex]{0pt}{3.5ex} CHaMP & HCO+ & 16 & 1.6  & MALT90 & C2H & 2012 & 23  \\ \hline
\rule[-1ex]{0pt}{3.5ex} HOPS & H2O & 11 & 14  & MALT90 & CH3CN & 2012 & 23  \\ \hline
\rule[-1ex]{0pt}{3.5ex} HOPS & NH3 (1-1) & 11 & 5.3  & MALT90 & H13CO+ & 2012 & 23  \\ \hline
\rule[-1ex]{0pt}{3.5ex} HOPS & NH3 (2-2) & 11 & 5.3  & MALT90 & H41alpha & 2012 & 23  \\ \hline
\rule[-1ex]{0pt}{3.5ex} FCRAO\_GRS & 13CO & 42 & 11  & MALT90 & HC13CCN & 2012 & 23  \\ \hline
\rule[-1ex]{0pt}{3.5ex} ThrUMMS & 12CO & 23 & 13  & MALT90 & HC3N & 2012 & 23  \\ \hline
\rule[-1ex]{0pt}{3.5ex} ThrUMMS & 13CO & 22 & 11  & MALT90 & HN13C & 2012 & 23  \\ \hline
\rule[-1ex]{0pt}{3.5ex} ThrUMMS & C18O & 23 & 11  & MALT90 & HNCO404 & 2012 & 23  \\ \hline
\rule[-1ex]{0pt}{3.5ex} ThrUMMS & CN & 23 & 12  & MALT90 & HNCO413 & 2012 & 23  \\ \hline
\rule[-1ex]{0pt}{3.5ex} NANTEN & 12CO & 2 & 1.1  & MALT90 & SiO & 2012 & 23  \\ \hline
\rule[-1ex]{0pt}{3.5ex} OGS & 12CO & 4 & 14  & VGPS & HI & 13 & 5.7  \\ \hline
\rule[-1ex]{0pt}{3.5ex} OGS & 13CO & 3 & 11  & CGPS & HI & 84 & 45  \\ \hline
\rule[-1ex]{0pt}{3.5ex} JCMT-HARP & 12CO & 92 & 24  & SGPS & HI & 13 & 4.4  \\ \hline
\end{tabular}
\end{center}
\end{table} 

\begin{table}[!ht]
\caption{VIALACTEA VLKB stored data collections consisting of 2D radio continuum images.} 
\label{tab:images}
\begin{center}       
\begin{tabular}{|l|l|r|r|}
\hline
\rule[-1ex]{0pt}{3.5ex} \textbf{Name} & \textbf{sub-survey} & \textbf{\# files} & \textbf{size [GB]} \\ \hline
\rule[-1ex]{0pt}{3.5ex} CORNISH & 5 GHz & 1408 & 84  \\ \hline
\rule[-1ex]{0pt}{3.5ex} MAGPIS & 1.4GHz & 352 & 1.4  \\ \hline
\rule[-1ex]{0pt}{3.5ex} Hi-Gal & 70$\mu$m & 166 & 7.2  \\ \hline
\rule[-1ex]{0pt}{3.5ex} Hi-Gal & 160$\mu$m & 166 & 3.7  \\ \hline
\rule[-1ex]{0pt}{3.5ex} Hi-Gal & 250$\mu$m & 166 & 2.2  \\ \hline
\rule[-1ex]{0pt}{3.5ex} Hi-Gal & 350$\mu$m & 166 & 1.3  \\ \hline
\rule[-1ex]{0pt}{3.5ex} Hi-Gal & 500$\mu$m & 166 & 0.6  \\ \hline
\rule[-1ex]{0pt}{3.5ex} MIPSGAL & 24$\mu$m & 339 & 13  \\ \hline
\rule[-1ex]{0pt}{3.5ex} WISE & 3.4$\mu$m & 694 & 44  \\ \hline
\rule[-1ex]{0pt}{3.5ex} WISE & 4.6$\mu$m & 694 & 44  \\ \hline
\rule[-1ex]{0pt}{3.5ex} WISE & 12$\mu$m & 694 & 44  \\ \hline
\rule[-1ex]{0pt}{3.5ex} WISE & 22$\mu$m & 694 & 44  \\ \hline
\end{tabular}
\end{center}
\end{table} 

\begin{table}[!ht]
\caption{VIALACTEA VLKB stored data collections.} 
\label{tab:extmaps}
\begin{center}       
\begin{tabular}{|l|l|r|r|l|}
\hline
\rule[-1ex]{0pt}{3.5ex} \textbf{Name} & \textbf{sub-survey} & \textbf{\# files} & \textbf{size [MB]} \\ \hline
\rule[-1ex]{0pt}{3.5ex} Extinction Maps & 5 arcmin resolution & 72 & 76  \\ \hline
\rule[-1ex]{0pt}{3.5ex} Extinction Maps & 10 arcmin resolution & 72 & 18 \\ \hline
\end{tabular}
\end{center}
\end{table} 

Hi-Gal data, the survey tiles listed in Table~\ref{tab:images} as well all the data derived from them (e.g. 
filaments, bubbles and single band catalogue listed in Table~\ref{tab:vlkbdb}) are private to the project.
MOPRA, GRS, NANTEN, OGS have been granted for usage inside the VIALACTEA project, but are covered by privacy 
policy.
Extinction Map data are private until the end of the project because they have been produced within the 
VIALACTEA project itself.
All the other data cube surveys and pointed archive listed in Table~\ref{tab:cubes} and \ref{tab:images}, are 
publicly available starting from locations listed in Appendix~\ref{app:data}.

All of the above files, that sum up to about 1 TB and nearly 40 000 files, are of course only the 
archival resource that needs to be made discoverable, accessible and retrievable to the community to enable 
the workflows and analysis that are the goal of the project.

The first, obvious, step in this direction is the ingestion of the needed metadata in a database back end to 
allow faster discovery and endpoint retrieval for accessibility. This is part of the content of the next 
subsection.

\subsection{Database Content}
\label{subsec:database}
Alongside the FITS data collections, a relational database (RDB) completes the VLKB resource content in terms 
of data information. This RDB (using a MySQL RDBMS) contains various types and blocks of information needed 
by the project. To mimic the logic underlying the project work packages subdivision, a set of schemata has 
been prepared to host the information. 

\begin{description}
		\item[Filaments \& Bubbles.] This database schema holds all the information related to the diffuse 
	objects identified from Hi-Gal continuum tiles (for filaments) and Hi-Gal and CORNISH tiles (for bubbles).
	
	Filamentary structures are described using a set of 3 tables, identifying \textit{filaments} as the 
	primary object, \textit{branches} as the ``linear'' components within them and their \textit{spines}, and 
	\textit{nodes} as the connection points of the various branch segments that compose a filament.
	
	Bubbles are described as a unique catalogue of diffuse objectS, because their roundish shape doesn't 
	require further relationships among their components.
	
	Both filaments and bubbles tables are completed with positional and global details plus contour 
	information. This latter is represented in two ways: an ordered sequence of sky positions (i.e. a polygon 
	outlining the diffuse structure) and a reference to a FITS file in MOC format (Multi-Order Coverage Map
	\cite{2014ivoa.spec.0602F}, an IVOA Recommendation), that is an HEALPix\cite{2005ApJ...622..759G}
	tessellation of the diffuse object's area to be used for easier cross-match with other positional features 
	in the VLKB or other databases.
	
		\item[Compact Sources.] This schema contains all the single band catalogues used by the project to 
	build up a \textit{bandmerged} catalogue of compact sources. The primary input of the band merge task 
	are the 5 single band catalogues derived from the Hi-Gal images through the CuTEx\cite{2011AA...530A.
	133M} tool (refined as a deliverable of the VIALACTEA project). 
	
	Alongside this catalogues, other have been 	
	used to build up a multi band catalogue of the galactic compact sources. The full single band catalogue 
	listing includes: Hi-Gal (PACS and SPIR bands), ATLASGAL, BGPS, MIPSGAL, MSX and WISE (bands W3 and W4) 
	catalogues for the galactic plane. The \textit{bandmerged} catalogue is produced from these ones using 
	data mining techniques included in the QFullTree tool delivered by the VIALACTEA project itself.
	
	This schema includes also other content, not used in the band-merging effort of the project, but used in 
	the SED scenario investigated by the project. One table is devoted to the grid of synthetic protocluster
	evolutionary	models (based on Robitaille\cite{2006ApJS..167..256R} models and delivered through the 
	project); this sums up to 20 million SED records to synthesize energy distributions in the bands used by 
	the band-merging effort. Another one is meant to keep track of the velocity information processed using 
	the radio cubes FITS and database through the Peak Finder tool developed by VIALACTEA WP3 members to be 
	later used in the distance estimation process of the various sources on the galactic plane.
	
		\item[Radio Cubes.] All the metadata information about the stored FITS files and needed for the search 
	and access interfaces is contained in this schema. This means, currently, all the sky frame boundaries 
	needed for the discovery phase plus the actual location in the storage for data processing (cutout and 
	merge) and retrieval.
	
	Alongside the datasets minimal descriptions this schema includes also a table with a the description of 
	all the sub-surveys to let the consumer applications (currently the Visual Analytics one developed within 
	the project) identify the relevant sub-survey for the user.
	
	This is probably the schema that may see more changes in the last phases of the project, if resources 
	allow to move to a better VO compliant scenario. In that case probably an 
	ObsCore\cite{2011ivoa.spec.1028T} table may take the role of part of the current FITS datasets 
	meta-descriptions.
	
		\item[TAP\_SCHEMA.] A last schema is kept in the VLKB to describe all of the schemata, tables, columns 
	and their keys and indexes to be deployed through the TAP interface. This is the schema that is part of 
	the IVOA TAP specification and acts as a counterpart to the information schema that is built in the RDBMS 	
	systems. Every change in the exposed database architecture that the VLKB wants to make visible to the user 
	as a description counterpart in this schema.
\end{description}

Table~\ref{tab:vlkbdb} summarizes the various catalogues ingested in the VLKB database part 
including the number of records. Not all of the database table details have been included, because this is highly dependent on the usage and, given the project is still ongoing, they will vary a lot.

\begin{table}[!ht]
\caption{VIALACTEA VLKB catalogue figures (rounded). Please note that figures may be partial or may differ from the final VLKB version because work is still in progress on the project at the time of this writing.} 
\label{tab:vlkbdb}
\begin{center}       
\begin{tabular}{|l|r|}
\hline
\rule[-1ex]{0pt}{3.5ex} \textbf{Catalogue} & \textbf{\# records} \\ \hline
\rule[-1ex]{0pt}{3.5ex} Filaments & 30 K \\ \hline
\rule[-1ex]{0pt}{3.5ex} Filament Branches & 150 K \\ \hline
\rule[-1ex]{0pt}{3.5ex} Bubbles & 5 K \\ \hline
\rule[-1ex]{0pt}{3.5ex} SED Models & 20 M \\ \hline
\rule[-1ex]{0pt}{3.5ex} WISE & 28 M \\ \hline
\rule[-1ex]{0pt}{3.5ex} ATLASGAL & 10 K \\ \hline
\rule[-1ex]{0pt}{3.5ex} BGPS & 8.6 K \\ \hline
\rule[-1ex]{0pt}{3.5ex} Hi-Gal (70 $\mu$m) & 160 K \\ \hline
\rule[-1ex]{0pt}{3.5ex} Hi-Gal (160 $\mu$m) & 600 K \\ \hline
\rule[-1ex]{0pt}{3.5ex} Hi-Gal (250 $\mu$m) & 470 K \\ \hline
\rule[-1ex]{0pt}{3.5ex} Hi-Gal (350 $\mu$m) & 250 K \\ \hline
\rule[-1ex]{0pt}{3.5ex} Hi-Gal (500 $\mu$m) & 130 K \\ \hline
\rule[-1ex]{0pt}{3.5ex} MIPSGAL & 2.6 M \\ \hline
\rule[-1ex]{0pt}{3.5ex} MSX & 400 K \\ \hline
\end{tabular}
\end{center}
\end{table} 

As for the datasets, some of these are publicly available, other are covered by a privacy policy.
Filaments and bubbles catalogues and SED models, being products of the project, are private until the end of 
the project itself. Hi-Gal derived data is part of the core of the project, and thus it is also covered by 
privacy policy. The other catalogues are public and available as listed in Appendix~\ref{app:data}.

Please note that, since the VIALACTEA project is not yet finished and delivered, the above figures and 
contents may not be exact, but anyway they give a good idea of the contents and goals of the VLKB.

\section{VLKB INTERFACES}
\label{sec:interfaces}
All the data and metadata contents described in Section~\ref{sec:collect} would be, of course, useless if not 
somehow discoverable and accessible. This is the goal of the interfaces that have been put on top of the 
database and storage system. The following subsections describe the two consuming interface systems that have 
been set up for the VIALACTEA community: a TAP interface (Sec.~\ref{subsec:tap}) and a set of dedicated 
interfaces (Sec.~\ref{subsec:dcm}) to consume to stored FITS datasets.

\subsection{TAP interface}
\label{subsec:tap}
The IVOA Table Access Protocol is a specification to allow generic table-sets to be deployed as VO resources. 
It aims at homogenizing database access in the astrophysical research field, adding specific metadata content 
for discoverability and interoperability among different resources.

The choice to have a TAP interface on top of the VLKB database part was made to allow both positional 
searches (that could have been done using a Simple Cone Search\cite{2008ivoa.specQ0222P} protocol) and 
generic searches through the fields of the ingested data, also considering cross correlations between 
multiple tables available through the service.

The TAP service has been deployed using the IA2 TAP implementation\footnote{
IA2TAP is part of the IA2 Data Access Layer service generation tools: 
\url{http://www.ia2.inaf.it/vo-services/vo-dance}}. It runs inside a GlassFish JAVA EE web container and it 
has been developed on top of the openCADC\footnote{https://github.com/opencadc} libraries by the Canadian 
Astronomy Data Center\footnote{http://www.cadc-ccda.hia-iha.nrc-cnrc.gc.ca/en/}.

Since the data ingested in the database is partly public and partly private data, the access to the interface 
has been secured. The authentication mechanism is currently being re-shaped to allow single user access (as 
opposite to the current project membership credentials), also in view of the user space related to the output 
products of the cutout and merge services described hereafter.

The TAP service, in short, serves everything included in the filaments, bubbles, compact sources, sed\_models 
and velocity peak store resource span and, in the future, it is intended to support also the radio cubes and 
images metadata. The visualization tool developed within the VIALACTEA project consumes this TAP service to 
provide the user with the relevant information to display alongside images and cubes.

\subsection{Discovery, Cutout \& Merge interfaces}
\label{subsec:dcm}
While the TAP interface is useful for generic tabular access, it is less powerful (at least alone) to perform 
positional discovery and is not designed to provide direct access to data files (unless one is willing only 
to retrieve a full file).

However, the requirements of the project were also to be able to:
\begin{itemize}
	\item identify what files' content overlap a region along a specific line of sight on the celestial sphere 
	and described by a circle or a rectangular range around it;
	\item cutout the files on positional constraints and also on the velocity axis (if 3D cubes were 
	investigated) to allow for efficient data transport on the net and work only on the relevant part of data;
	\item merge adjacent files (from the same sub-survey, i.e. no information post-processing during the merge 
	phase) into a single image/cube based on positional and velocity bounds.
\end{itemize}

IVOA specifications are currently covering the second point (see SODA\footnote{SODA Proposed Recommendation 
\url{http://www.ivoa.net/documents/SODA/index.html}}), were in the phase of describing the first one when the 
VIALACTEA project was kicked-off (the SIAv2\cite{2015ivoa.spec.1223D} protocol) and foresee the third one in 
the near future (probably a minor revision of SODA).

Thus, even if keeping an eye on what IVOA was providing, a custom solution was undertook to have the 
discovery and access interfaces for VIALACTEA in place. This custom solution allowed also to work directly in 
galactic coordinates, while in the IVOA interfaces equatorial coordinates are prioritized.

The developed interface architecture is now composed of three services that share the same parametric query 
solution for positions, bound descriptions, sub-survey selection and are able to search and cutout
indifferently of the underlying FITS header metadata (at least as far the WCS\cite{2002AA...395.1077C,2006AA...446..747G} descriptors are correctly in 
place and understood by the AST\footnote{http://starlink.eao.hawaii.edu/starlink/AST} libraries).
Specifically, the headers of various sub-surveys are passed to
 AST\cite{2016A&C....15...33B} library which is used to perform region overlap,
and boundary computations. The AST library is a C-library so the JNI technology is used to access 
C-code from Java domain. Most of the Java-code is performing interfacing. Essential work is done "below" JNI in C. 
We process the partial results from AST in C-domain and then return the final results to Java domain. 
So we cross the JNI once per request.

Similar approach is taken also for the merge service, where the used engine is 
Montage\footnote{http://montage.ipac.caltech.edu/}. The merging service internally adds 3D-datacubes 
which are the result of the cutout service.

Issues on header metadata content as been solved, as far as possible, through corrected information ingestion 
at database level, trying not to change the original retrieved FITS, especially in the case of mirroring pre-
existing resources. Data content of the FITS files also has been kept to a minimum, untouched in the cutout 
service, at most regridded along the third axis in the merge service if the adjacent FITS had different 
binning on that axis.

The set of three services, secured like the TAP interface to prevent unauthorized access to private data, 
deploys to the project's members all the images, cubes and extinction maps summarized in Tables~
\ref{tab:cubes}, \ref{tab:images} and \ref{tab:extmaps}. The VIALACTEA visualization tool as well as other 
task tools make use of these services endpoints to prepare the input data for their tasks.

\section{PLANNED VLKB SETUP}
\label{sec:status}

\begin{figure}[!ht]
\begin{center}
\begin{tabular}{c} 
\includegraphics[width=0.7\textwidth]{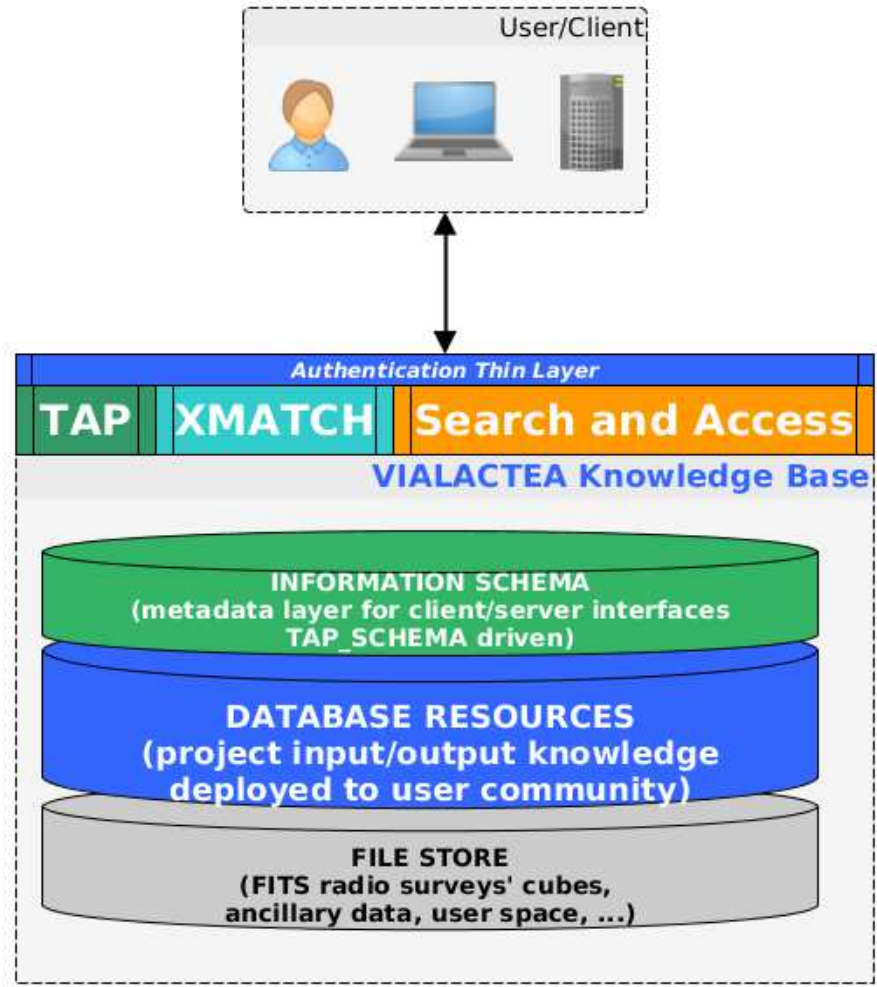}
\end{tabular}
\end{center}
\caption[] 
{ \label{fig:vlkboverview} 
VLKB content and interfaces overview. Shown, on the interfaces layer, the positional cross match service based on HEALPix tessellation, not yet fully developed.}
\end{figure} 

The data resources described in Section~\ref{sec:collect} and the interfaces reported in the previous 
Section~\ref{sec:interfaces} build up what was the goal of the VIALACTEA Work Package 5 Task 1.
Currently the development is in finalization, but all what as been reported is planned to be ready and 
consumable by the end of September 2016 (i.e. the end of the funded project). Figure~\ref{fig:vlkboverview} 
gives a quick view of the architecture of the full system.

Depending mainly on time resource availability, more efficient discovery and access solutions, based on a 
modular and distributed architecture, may be set in place and a better VO compliance can be achieved.

One major point, not detailed in this paper, but planned to be ready by the end of the project, is the 
improvement of positional searches with the usage of HEALPix tessellation, in particular a dedicated service 
should be set up to allow direct, quick, cross match on positions between diffuse objects (filaments and 
bubbles) and compact sources. Given the large amount of records in the DB, this would be a welcome addition 
to the available interfaces.

Also planned is the change from a membership-wide user to a single-based user authentication. This as the 
goal also to help maintain the staging area for the cutout and merged FITS outputs of the access 
interfaces, that otherwise may grow indefinitely and with no choice, for the user, to retrieve a product 
already calculated.

\section{CONCLUSIONS}
\label{sec:close}
Building a data infrastructure to cover large part of the radio data for the galactic plane has been quite a 
big challenge, especially if undertaken with minimal resources and trying to cope with the short schedule of 
a three-year project with frequently updated requirements.

In this view the VLKB development has profited from the knowledge and experience available at the INAF IA2 
data center, not last the VO experience that was highly useful in defining the solutions for discovery and 
access to the data, even if not directly available in terms of IVOA Recommendations.

One main critic that may be addressed to the solutions put in place is that large part of the data has been 
mirrored in a single location to be again deployed to the community. Indeed, initially the idea was to build 
as much as possible over existing resources and services; however most of the data re-used fell behind custom 
interfaces, usually graphical web forms, that were difficult to access programmatically and attach to a 
common interface solution.

Considering that the VIALACTEA project is now running to its final deadline, it is difficult to foresee what 
improvements may be made to the VLKB, besides minimal maintenance; however, if further development and 
progress were to be foreseen, for sure developing a real VO compliant architecture will be definitely of a 
benefit to the community not to have future projects develop again mirrored interfaces to already existing 
resources.

\appendix

\section{PUBLIC DATA}
\label{app:data}
Here follows the listing of all the public data used in developing the VLKB content.

\subsection{FITS cubes}
Radio cubes surveys and pointed archives.
\begin{description}
	\item[CHIMP:] \url{http://dx.doi.org/10.11570/16.0001} 
	\item[CHaMP:] \url{http://www.astro.ufl.edu/~peterb/research/champ/rbank/}
	\item[HOPS:] \url{http://awalsh.ivec.org/hops/public/data_cubes.php}
	\item[ThrUMMS:] \url{http://www.astro.ufl.edu/~peterb/research/thrumms/rbank/}
	\item[JCMT-HARPS:] \url{http://www.cadc-ccda.hia-iha.nrc-cnrc.gc.ca/en/jcmt/}
	\item[MALT90:] \url{http://atoa.atnf.csiro.au/MALT90}
	\item[VGPS:] \url{http://www.ras.ucalgary.ca/VGPS/}
	\item[CGPS:] \url{http://www.ras.ucalgary.ca/CGPS/products/}
	\item[SPGS:] \url{http://www.atnf.csiro.au/research/HI/sgps/queryForm.html}
\end{description}

\subsection{FITS images}
Continuum surveys.
\begin{description}
	\item[CORNISH:] \url{http://cornish.leeds.ac.uk/public/index.php}
	\item[MAGPIS:] \url{http://third.ucllnl.org/gps/}
	\item[MIPSGAL:] \url{http://irsa.ipac.caltech.edu/data/SPITZER/MIPSGAL/}
	\item[WISE:] \url{http://irsa.ipac.caltech.edu/Missions/wise.html}
\end{description}

\subsection{Catalogues}
Single band catalogues used in band-merging.
\begin{description}
	\item[WISE:] \url{http://irsa.ipac.caltech.edu/cgi-bin/Gator/nph-dd}
	\item[ATLASGAL:] \url{http://atlasgal.mpifr-bonn.mpg.de/cgi-bin/ATLASGAL_DATABASE.cgi}
	\item[BGPS:] \url{http://irsa.ipac.caltech.edu/cgi-bin/Gator/nph-scan?submit=Select&projshort=BOLOCAM}
	\item[MIPSGAL:] \url{http://vizier.u-strasbg.fr/viz-bin/VizieR?-source=J/AJ/149/64}
	\item[MSX:] \url{http://irsa.ipac.caltech.edu/applications/MSX/MSX/mission.htm}
\end{description}

\acknowledgments
This work has been developed under the European Community 7th Framework
Programme, Grant Agreement 607380, VIALACTEA - The Milky Way as a Star formation
engine.

The VLKB resource data and metadata are highly depending on external public and private sources. We 
acknowledge the efforts of the staff of the primary repositories and archives from which we retrieved 
the data to be offered to the VIALACTEA members and, at the end of the project, to the astrophysical 
community (all the relevant starting points are listed in Appendix~\ref{app:data}).

This research made use of Montage. It is funded by the National Science Foundation under Grant Number 
ACI-1440620, and was previously funded by the National Aeronautics and Space Administration's Earth 
Science Technology Office, Computation Technologies Project, under Cooperative Agreement Number 
NCC5-626 between NASA and the California Institute of Technology.

\bibliography{MMolinaro-SPIE-ATeI-2016-9913-17} 

\begin{thebibliography}{10}

\bibitem{2010ivoa.spec.0327D}
{Dowler}, P., {Rixon}, G., and {Tody}, D., ``{Table Access Protocol Version
  1.0}.'' IVOA Recommendation 27 March 2010 (Mar. 2010).

\bibitem{2010AA...524A..42P}
{Pence}, W.~D., {Chiappetti}, L., {Page}, C.~G., {Shaw}, R.~A., and {Stobie},
  E., ``{Definition of the Flexible Image Transport System (FITS), version
  3.0},'' {\em Astronomy and Astrophysics}~{\bf 524},  A42 (Dec. 2010).

\bibitem{2014ivoa.spec.0602F}
{Fernique}, P., {Boch}, T., {Donaldson}, T., {Durand}, D., {O'Mullane}, W.,
  {Reinecke}, M., and {Taylor}, M., ``{MOC - HEALPix Multi-Order Coverage map
  Version 1.0}.'' IVOA Recommendation 02 June 2014 (June 2014).

\bibitem{2005ApJ...622..759G}
{G{\'o}rski}, K.~M., {Hivon}, E., {Banday}, A.~J., {Wandelt}, B.~D., {Hansen},
  F.~K., {Reinecke}, M., and {Bartelmann}, M., ``{HEALPix: A Framework for
  High-Resolution Discretization and Fast Analysis of Data Distributed on the
  Sphere},'' {\em Astrophysical Journal}~{\bf 622},  759--771 (Apr. 2005).

\bibitem{2011AA...530A.133M}
{Molinari}, S., {Schisano}, E., {Faustini}, F., {Pestalozzi}, M., {di Giorgio},
  A.~M., and {Liu}, S., ``{Source extraction and photometry for the
  far-infrared and sub-millimeter continuum in the presence of complex
  backgrounds},'' {\em Astronomy and Astrophysics}~{\bf 530},  A133 (June
  2011).

\bibitem{2006ApJS..167..256R}
{Robitaille}, T.~P., {Whitney}, B.~A., {Indebetouw}, R., {Wood}, K., and
  {Denzmore}, P., ``{Interpreting Spectral Energy Distributions from Young
  Stellar Objects. I. A Grid of 200,000 YSO Model SEDs},'' {\em Astrophysical
  Journal Suppl.}~{\bf 167},  256--285 (Dec. 2006).

\bibitem{2011ivoa.spec.1028T}
{Tody}, D., {Micol}, A., {Durand}, D., {Louys}, M., {Bonnarel}, F., {Schade},
  D., {Dowler}, P., {Michel}, L., {Salgado}, J., {Chilingarian}, I., {Rino},
  B., {de Dios Santander}, J., and {Skoda}, P., ``{Observation Data Model Core
  Components, its Implementation in the Table Access Protocol Version 1.0}.''
  IVOA Recommendation 28 October 2011 (Oct. 2011).

\bibitem{2008ivoa.specQ0222P}
{Plante}, R., {Williams}, R., {Hanisch}, R., and {Szalay}, A., ``{Simple Cone
  Search Version 1.03}.'' IVOA Recommendation 22 February 2008 (Feb. 2008).

\bibitem{2015ivoa.spec.1223D}
{Dowler}, P., {Bonnarel}, F., and {Tody}, D., ``{IVOA Simple Image Access
  Version 2.0}.'' IVOA Recommendation 23 December 2015 (Dec. 2015).

\bibitem{2002AA...395.1077C}
{Calabretta}, M.~R. and {Greisen}, E.~W., ``{Representations of celestial
  coordinates in FITS},'' {\em Astronomy and Astrophysics}~{\bf 395},
  1077--1122 (Dec. 2002).

\bibitem{2006AA...446..747G}
{Greisen}, E.~W., {Calabretta}, M.~R., {Valdes}, F.~G., and {Allen}, S.~L.,
  ``{Representations of spectral coordinates in FITS},'' {\em Astronomy and
  Astrophysics}~{\bf 446},  747--771 (Feb. 2006).

\bibitem{2016A&C....15...33B}
{Berry}, D.~S., {Warren-Smith}, R.~F., and {Jenness}, T., ``{AST: A library for
  modelling and manipulating coordinate systems},'' {\em Astronomy and
  Computing}~{\bf 15},  33--49 (Apr. 2016).

\end{thebibliography}
\bibliographystyle{spiebib} 

\end{document}